\documentclass[nofootinbib,amsmath,amsfonts]{revtex4}
\usepackage{hyperref}
\usepackage{longtable}
\usepackage{graphicx,xcolor}

\newcommand{\eqr}[1]{Eq.~\eqref{#1}}

\newcommand{\ssecr}[1]{Subsec.~[\ref{#1}]}
\newcommand{\secr}[1]{Sec.~[\ref{#1}]}
\newcommand{\appr}[1]{Appendix~[\ref{#1}]}
\newcommand{\measure}{{\rm{m}}}

\newcommand{\vR}{{\mathbf{r}}}

\newcommand{\Kappa}{{\cal K}}
\newcommand{\nrange}{\overline{1,n}}
\newcommand{\nmorange}{\overline{1,n\!-\!1}}
\newcommand{\arglist}[2]{#1_{1}, \, \ldots, \, #1_{#2}}
\newcommand{\arglistr}[2]{#1_{1}(\vR), \, \ldots, \, #1_{#2}(\vR)}

\newcommand{\alTcxi}{T, \, c, \, \xi}

\newcommand{\spd}{\!\cdot\!}
\newcommand{\PCDF}[2]{{\frac{\partial #1}{\partial x_{#2}}}} 
\newcommand{\SumVarea}{\vphantom{\sum\limits_1^1}}
\newcommand{\BigVarea}{\vphantom{\Big(\Big)}}

\newcommand{\vn}{{\mathbf{n}}}

\long\def\reffootnote[#1]#2{\begingroup\def\thefootnote{\fnsymbol{footnote}}\footnote[#1]{#2}\endgroup}
\newcommand{\tsup}[1]{\textsuperscript{#1}}

\graphicspath{{figures/}}

\begin{document}

\title{The square gradient model in a two-phase mixture I. \\Equilibrium properties.}
\author{K.~S.~Glavatskiy}
\author{D.~Bedeaux}
\affiliation{Department of Chemistry, Norwegian University of Science and Technology, Trondheim, Norway}
\date\today

\begin{abstract}
In order to describe a nonuniform equilibrium mixture with an interface between two coexisting phases it is necessary to consider contributions to the Helmholtz energy which depend on the
gradients of for instance the density. Van der Waals \cite{vdW/sg, vdW/translation} was the first to introduce such a term, which is very important in the interfacial region for a one-component
system. Cahn \& Hilliard \cite{cahnhilliard/fens/I} extended this analysis to a binary mixture by introducing gradient terms of the mol fraction. We give an systematic extension of the gradient
theory to three-dimensional multi-component systems.
\end{abstract}

\maketitle
\numberwithin{equation}{section}


\section{Introduction.}

In order to describe the equilibrium properties of an interface between two coexisting phases, using a continuous model, it is necessary to consider contributions to the Helmholtz energy which
depend on the gradients of for instance the density \cite{RowlinsonWidom}. Van der Waals \cite{vdW/sg, vdW/translation} was in 1893 the first to introduce such a term for a one-component system.
In 1958 Cahn \& Hilliard \cite{cahnhilliard/fens/I} extended the analysis of van der Waals and introduced gradient terms of the mol fraction in binary mixtures. As the Helmholtz energy density
given by van der Waals is no longer a function of the local density or local densities alone, there is no local equilibrium in the traditional sense in the interfacial region. The continuous
description is in other words "not autonomous". We refer to the monograph by Rowlinson and Widom \cite{RowlinsonWidom} for a thorough discussion of the van der Waals model in general and of this
point in particular. A lot of work on the equilibrium gradient model was done by Cornelisse \cite{cornelisse/grad}. We refer to his thesis for the relevant references.

The gradient model is often used for a system, in which properties vary only in one direction. We do not restrict ourself in this manner and do the analysis for the three-dimensional system.

We use the standard thermodynamic variables which obey the standard thermodynamic relations for homogeneous mixtures. In \secr{sec/NonUniform} we consider the inhomogeneous mixture. We postulate
the dependence of the specific Helmholtz energy on the thermodynamic variables and their gradients, using the fact, that in equilibrium the temperature is constant. Using that in equilibrium the
total Helmholtz energy has a minimum and that the amount of the various components is fixed, expressions are derived for the chemical potentials of all the components, using Lagrange's method.
This is done for a different choice of variables in \secr{sec/NonUniform/Matter} and \secr{sec/NonUniform/Volume}. Extending a method developed by Yang et. al. \cite{Yang/surface} an expression
is also found for the pressure tensor. We derive the different forms of the Gibbs relations. Their importance is crucial for non-equilibrium description which should be based on the equilibrium
analysis. Explicit expressions are also given for the internal energy, enthalpy and Gibbs energy densities. In \secr{sec/Discussion} we give a discussion and conclusion.

In following papers we aim to generalize the analysis to non-equilibrium systems, using the results of this paper. This will extend the work of Bedeaux et. al. \cite{bedeaux/vdW/I,
bedeaux/vdW/II, bedeaux/vdW/III} for one-component systems, in which the properties varied only in one direction.

\section{The gradient model.}\label{sec/NonUniform}

In order to describe inhomogeneous systems in equilibrium, one could assume, that this can be done by the usual thermodynamic variables, which depend on the spatial coordinates. All standard
thermodynamic relations are then assumed to remain valid.

As van der Waals has shown for one-component system \cite{vdW/sg}, however, this is not enough to describe the surface. It is necessary to assume that thermodynamic potentials, particularly the
Helmholtz energy density, also depends on spatial derivatives of the density. Cahn \& Hilliard have shown for a binary mixture \cite{cahnhilliard/fens/I}, that the Helmholtz energy should depend
on the gradients of the mole fraction of one of the components. For a multi-component non-polarizable mixture we shall use as general form of the Helmholtz energy
\begin{equation}\label{eq/NonUniform/01}
f(\vR) = f_{0}(T;\,\arglistr{z}{n}) + {\frac{1}{2}}{\sum\limits_{i,j=1}^{n}{\widetilde\kappa_{ij}(\arglistr{z}{n})\,\nabla{z_{i}(\vR)}\spd\nabla{z_{j}(\vR)}}}
\end{equation}
where $f_{0}$ is the homogeneous Helmholtz energy and $z_{k},\, k=\nrange$, where $\nrange$ indicates all integers from $1$ to $n$, are generalized densities. These can be either total
concentration of the mixture together with the $n-1$ fractions of the components, $\{c,\,\arglist{\xi}{n-1}\}$, (see \secr{sec/NonUniform/Matter}) or $n$ concentrations of the components,
$\{\arglist{c}{n}\}$, (see \secr{sec/NonUniform/Matter}). Concentrations may be taken either on the molar basis or on the mass basis.

This form of the specific Helmholtz energy is obtained by the following arguments (cf. also \cite{vdW/translation, cahnhilliard/fens/I, Yang/surface}). Dependence of the specific Helmholtz
energy on the density gradients can be represented by Taylor series in these gradients. We describe here isotropic fluids, so any coefficient in this Taylor series can not depend on any
direction and thus should be scalar. The zeroth term, taken when all the gradients are equal to zero, is the homogeneous Helmholtz energy $f_{0}(T;\,\arglistr{z}{n})$. In equilibrium, the total
Helmholtz energy of the system has a minimum. Thus, the first order term, with the first order density gradients, $\nabla{z_{i}(\vR)}$, is zero. The second order term is quadratic in the first
order density gradients $\nabla{z_{i}(\vR)}\spd\nabla{z_{j}(\vR)}$, and is linear in the second order density gradients $\nabla^{2}{z_{i}(\vR)}$. The latter one, however, contributes to the
total Helmholtz energy the same way as the former one:
\begin{equation}\label{eq/NonUniform/01a}
\begin{array}{ll}
\int_{V}d\vR\,\widetilde{\kappa}_{i}\,\nabla^{2}{z_{i}(\vR)} &= %
- \int_{V}d\vR\,\nabla{\widetilde{\kappa}_{i}}\spd\,\nabla{z_{i}(\vR)} + \int_{V}d\vR\,\nabla\spd(\widetilde{\kappa}_{i}\,\nabla{z_{i}(\vR)})%
= \\ \\ &= %
-\int_{V}d\vR\,{\sum\limits_{j=1}^{n}{{\frac{\partial\widetilde{\kappa}_{i}}{\partial z_{j}}}\,\nabla{z_{j}(\vR)}\spd\nabla{z_{i}(\vR)}}} +
\int_{S}dS\,\widetilde{\kappa}_{i}\,\vn_{s}\spd\nabla{z_{i}(\vR)}
\end{array}
\end{equation}
The first term on the right hand side of this equation can be combined with the quadratic term in the first order gradients. The second one can be chosen equal to zero by proper choice of the
boundaries of integration. Thus, we end up with \eqr{eq/NonUniform/01} for the specific Helmholtz energy where the coefficients $\widetilde\kappa_{ij}$ are the combinations of those from the
quadratic in the first order density gradients term and the corresponding ones from the linear in the second order density gradients term. For ease of notation we will write
$\widetilde\kappa_{ij}$ instead of $\widetilde\kappa_{ij}(\arglistr{\widetilde z}{n})$, remembering their dependence on these variables.

We note the ambiguity in the definition of the specific Helmholtz energy. Different expressions for $f(\vR)$ give the same expression for the total Helmholtz energy $F$, due to the cancellation
of the boundary contributions. This can be interpreted such that the measurable quantity is only the total Helmholtz energy, but not the specific one. To build the local description we need the
local quantities, however. We will use \eqr{eq/NonUniform/01} for the specific Helmholtz energy, remembering that a divergence of a vector field, the normal component of which is zero on the
boundary, can in principal be added. We will return to this point in the \appr{sec/Ambiguity}.

We may choose the matrix $\widetilde\kappa_{ij}$ to be symmetric with respect to the component number ($\widetilde\kappa_{ij} = \widetilde\kappa_{ji}$) without loss of generality (since it
appears only in symmetric combinations). We shall always take $\widetilde\kappa_{ij}$ independent of the temperature.

We note, that the gradient model is, as it has been used here, a general approach and is not only applied to surfaces. It has, for instance, been used in the description of critical behavior using
renormalization group theory \cite{GL, wilson/rgcp}. In this paper we will focus on its use for the description of the surface.

In the following paper, where we extend the analysis to non-equilibrium two-phase mixtures, we need all the thermodynamic variables. We derive all thermodynamic quantities and relations for the
given choice of the independent variables. This is done in \secr{sec/NonUniform/Matter} for molar specific variables. We determine how the Helmholtz energy varies with a change of the variables
and with a change of position and obtain so-called Gibbs relations. In \ssecr{sec/NonUniform/Matter/Surface} we determine the physical meaning of the Lagrange multipliers and other quantities,
for which expressions were derived. In \secr{sec/NonUniform/Volume} we derive the results for volume specific variables following the same procedure as in \secr{sec/NonUniform/Matter}.

\section{Gradient model for the molar variables.}\label{sec/NonUniform/Matter}
\subsection{The Lagrange method.}\label{sec/NonUniform/Matter/Lagrange}

We write the Helmholtz energy as
\begin{equation}\label{eq/NonUniform/03}
f(\vR) = f_{0}(T,\,c,\,\xi) + \Kappa(c,\,\xi,\,\nabla{c},\,\nabla{\xi})
\end{equation}
where
\begin{equation} \label{eq/NonUniform/08a}
\Kappa(c,\,\xi,\,\nabla{c},\,\nabla{\xi}) \equiv  {\frac{1}{2}}{\frac{\kappa}{c}}\,|\nabla{c}\,|^{2} + {\sum\limits_{i=1}^{n-1}{{\frac{\kappa_{i}}{c}}\,\nabla{c}\spd\nabla{\xi_{i}}}} +
{\frac{1}{2}}{\sum\limits_{i,j=1}^{n-1}{{\frac{\kappa_{ij}}{c}}\,\nabla{\xi_{i}}\spd\nabla{\xi_{j}}}}
\end{equation}
Here and further we suppress $\vR$ as an argument where this is not confusing. We also use $\xi$ as short notation instead of whole set $\{\arglist{\xi}{n-1}\}$ and $\nabla{\xi}$ as short notation
instead of $\{\arglist{\nabla{\xi}}{n-1}\}$. The molar density distributions are such that they minimize the total Helmholtz energy
\begin{equation}\label{eq/NonUniform/04}
F = \int_{V}d\vR\,c(\vR)\,f(\vR)
\end{equation}
Assuming that no chemical reactions occur, the total number of moles of each component, $\nu_{i} = \int_{V}d\vR\,\xi _{i}(\vR)\,c(\vR)$ for $i=\nmorange$, as well as the total number of moles,
$\nu = \int_{V}d\vR\,c(\vR)$, are constant. The problem of minimizing the functional \eqref{eq/NonUniform/04}, having $n$ constraints can be done using the Lagrange method. Thus we minimize the
integral
\begin{equation}\label{eq/NonUniform/06}
\Omega = \int_{V}d\vR\,c(\vR)\Big[ f(\vR) - \mu_{n} - {\sum\limits_{i=1}^{n-1}{\psi_{i}\,\xi_{i}(\vR)}} \Big]\equiv - \int_{V}d\vR\,p\,(\vR)
\end{equation}%
where $\mu _{n}$ and $\psi _{i}$ are scalar Lagrange multipliers. The concentration distributions which minimize the integral \eqref{eq/NonUniform/06} must be solutions of the corresponding
Euler-Lagrange equations. These relations give for the introduced Lagrange multipliers:
\begin{equation}\label{eq/NonUniform/08}
\begin{aligned}
\mu_{n} & =  {\frac{\partial}{\partial c}}\Big(c\,(f_{0} + \Kappa)\Big) - {\sum\limits_{i=1}^{n-1}{\psi_{i}\xi _{i}}} - \nabla\spd\Big(\kappa\,\nabla{c} + {\sum\limits_{i=1}^{n-1}{\kappa_{i}\,\nabla{\xi_{i}}}} \Big) \\%
\psi_{k} & = {\frac{\partial}{\partial \xi_{k}}}\Big(f_{0} + \Kappa\Big) - {\frac{1}{c}}\,\nabla\spd\Big(\kappa_{k}\,\nabla{c} + {\sum\limits_{i=1}^{n-1}{\kappa_{ik}\,\nabla{\xi_{i}}}}\Big) , \quad k=\nmorange %
\end{aligned}
\end{equation}%
and an expression for $p\,$:
\begin{equation} \label{eq/NonUniform/09}
p\,(\vR) = c^{2}{\frac{\partial}{\partial c}}\big(f_{0} + \Kappa\big) - c\,\nabla\spd\Big(\kappa\,\nabla{c} + {\sum\limits_{i=1}^{n-1}{\kappa_{i}\,\nabla{\xi_{i}}}}\Big)
\end{equation}

The ambiguity in $f(\vR)$ discussed above does not affect the expressions for $\mu_{n}$, $\psi_{k}$ and $p(\vR)$ (see \appr{sec/Ambiguity}). We will see in \ssecr{sec/NonUniform/Matter/Surface}
that the Lagrange multipliers $\mu_{n}$ and $\psi_{k}$ are the chemical potentials of the components and $p(\vR)$ is a pressure. The exact meaning of this pressure as well as the meaning of the
other quantities derived in this Subsection will be discussed in \ssecr{sec/NonUniform/Matter/Surface}.

Solving \eqr{eq/NonUniform/08} for $c$ and $\xi$, one obtains the density profiles for the system. To do this one needs the values for the Lagrange multipliers $\mu_{n}$ and $\psi_{k}$.

Multiplying the first of the equations in \eqr{eq/NonUniform/08} with $\nabla{c(\vR)}$ and the other $(n-1)$ ones with $\nabla{\xi_{k}(\vR)}$ and summing them all up\footnote{This method is a
generalization of the one, given by Yang et. al. \cite{Yang/surface} for a one-component system.} we obtain the following expression
\begin{equation}\label{eq/NonUniform/10}
{\frac{d\sigma_{\alpha\beta}(\vR)}{dx_{\alpha }}}=0
\end{equation}%
where we use the summation convention over double Greek subscripts. The tensor
\begin{equation}\label{eq/NonUniform/11}
\sigma_{\alpha\beta}(\vR) = p\,(\vR)\,\delta_{\alpha\beta} + \gamma_{\alpha\beta}(\vR)
\end{equation}%
will be identified as the pressure tensor. Furthermore the tensor
\begin{equation}\label{eq/NonUniform/12}
\gamma_{\alpha\beta}(\vR) = \kappa\,{\PCDF{c(\vR)}{\alpha}}{\PCDF{c(\vR)}{\beta}} + {\sum\limits_{i=1}^{n-1}{\kappa_{i}\Big({\PCDF{\xi_{i}(\vR)}{\alpha}}{\PCDF{c(\vR)}{\beta}} +
{\PCDF{c(\vR)}{\alpha}}{\PCDF{\xi_{i}(\vR)}{\beta}} \Big)}} + {\sum\limits_{i,j=1}^{n-1}{\kappa_{ij}\,{\PCDF{\xi_{i}(\vR)}{\alpha}}{\PCDF{\xi_{j}(\vR)}{\beta}} }}
\end{equation}%
will be referred to as the tension tensor\footnote{The explicit expression for the pressure tensor in the square gradient model for a multi-component mixture was, to the best of our knowledge,
not given before.}. We note, that both $\sigma_{\alpha\beta}(\vR)$ and $\gamma_{\alpha \beta }(\vR)$ are symmetric tensors.

From the definition \eqref{eq/NonUniform/06} of the $p(\vR)$, we can see that the Helmholtz energy given in \eqref{eq/NonUniform/03} and quantities which are given by the \eqr{eq/NonUniform/08}
and \eqr{eq/NonUniform/09} are related in the following way
\begin{equation}\label{eq/NonUniform/15}
f(\vR) = \mu_{n} + {\sum\limits_{i=1}^{n-1}{\psi_{i}\,\xi_{i}(\vR)}} - p\,(\vR)\,v(\vR)
\end{equation}%
Including a possible divergence term in $f(\vR)$ would modify this equation by adding this term also on the right hand side.

\subsection{Gibbs relations.}\label{sec/NonUniform/Matter/Gibbs}

Consider the variation of the total Helmholtz energy $\delta F[T,\,c(\vR),\,\xi(\vR),\,\nabla{c(\vR)},\,\nabla{\xi(\vR)}]$ with respect to the variation of the variables, which it depends on:
\begin{equation}\label{eq/NonUniform/50}
\delta F[T,\,c,\,\xi,\,\nabla{c},\,\nabla{\xi}] = \int_{V}d\vR\,\{f(T,\,c,\,\xi,\,\nabla{c},\,\nabla{\xi})\,\delta c + c\,\delta f(T,\,c,\,\xi,\,\nabla{c},\,\nabla{\xi})\}
\end{equation}
where
\begin{equation}\label{eq/NonUniform/51}
\delta f(T,\,c,\,\xi,\,\nabla{c},\,\nabla{\xi}) = {\frac{\partial f}{\partial T}}\,\delta T + {\frac{\partial f}{\partial c}}\,\delta c + {\sum\limits_{i=1}^{n-1}{{\frac{\partial
f}{\partial\xi_{i}}}\,\delta\xi_{i}}} + {\frac{\partial f}{\partial \nabla{c}}}\,\delta \nabla{c} + {\sum\limits_{i=1}^{n-1}{{\frac{\partial f}{\partial\nabla{\xi_{i}}}}\,\delta\nabla{\xi_{i}}}}
\end{equation}
is the total thermodynamic differential of the specific Helmholtz energy with respect to the thermodynamic variables, which it depends on. Given \eqr{eq/NonUniform/03}, \eqr{eq/NonUniform/08}
and \eqr{eq/NonUniform/09}, \eqr{eq/NonUniform/51} becomes
\begin{equation}\label{eq/NonUniform/52}
\delta f(T,\,c,\,\xi,\,\nabla{c},\,\nabla{\xi}) = {\frac{\partial f_{0}}{\partial T}}\,\delta T + {\frac{p}{c^{2}}}\,\delta c + {\sum\limits_{i=1}^{n-1}{\psi_{k}\,\delta\xi_{i}}} +
{\frac{1}{c}}\,\nabla\spd{\delta\mathbf\Theta}
\end{equation}
where
\begin{equation}\label{eq/NonUniform/53}
{\delta\mathbf\Theta}(c,\,\xi,\,\nabla{c},\,\nabla{\xi}) \equiv \Big(\kappa\,\nabla{c}+{\sum\limits_{i=1}^{n-1}{\kappa_{i}\,\nabla\xi_{i}}}\Big)\,\delta c +
{\sum\limits_{k=1}^{n-1}{\Big(\kappa_{k}\,\nabla{c}+{\sum\limits_{i=1}^{n-1}{\kappa_{ik}\,\nabla\xi_{i}}}\Big)\,\delta\xi_{k}}}
\end{equation}

The total Helmholtz energy variation becomes then
\begin{equation}\label{eq/NonUniform/54}
\delta F[T,\,c,\,\xi,\,\nabla{c},\,\nabla{\xi}] = \int_{V}d\vR\,\Big\{f(T,\,c,\,\xi,\,\nabla{c},\,\nabla{\xi})\,\delta c + %
c\,\Big({\frac{\partial f_{0}}{\partial T}}\,\delta T + {\frac{p}{c^{2}}}\,\delta c + {\sum\limits_{i=1}^{n-1}{\psi_{k}\,\delta\xi_{i}}}\Big)\Big\}
\end{equation}
since the boundary integral $\int_{S}dS\,\vn_{s}\spd{\delta\mathbf\Theta}$ disappears, because we have chosen boundaries of the system such that the density gradients are zero everywhere along
the boundaries. Thus, we will interpret the expression in parenthesis as the total thermodynamic differential of the specific Helmholtz energy:
\begin{equation}\label{eq/NonUniform/56}
\delta f(T,\,c,\,\xi,\,\nabla{c},\,\nabla{\xi}) = {\frac{\partial f_{0}}{\partial T}}\,\delta T + {\frac{p}{c^{2}}}\,\delta c + {\sum\limits_{i=1}^{n-1}{\psi_{k}\,\delta\xi_{i}}}
\end{equation}

We note the ambiguity in the definition of the total thermodynamic differential of the specific Helmholtz energy. Different expressions \eqr{eq/NonUniform/52} and \eqr{eq/NonUniform/56} for
$\delta f$ give the same expression \eqr{eq/NonUniform/54} for $\delta F$, due to the cancellation of the boundary contributions. This can be interpreted such that the measurable quantity is
only the total thermodynamic differential of the total Helmholtz energy, but not the total thermodynamic differential of the specific one. This ambiguity is similar to the ambiguity in
definition of the specific Helmholtz energy. We will use \eqr{eq/NonUniform/56}, remembering this ambiguity.

We write \eqr{eq/NonUniform/56} in the form
\begin{equation}\label{eq/NonUniform/34a}
\delta f(T, v(\vR), \arglistr{\xi}{n-1}) = -s(\vR)\,\delta T - p\,(\vR)\,\delta v(\vR) + {\sum\limits_{i=1}^{n-1}{\psi _{i}\,\delta \xi _{i}(\vR)}} %
\end{equation}%
where
\begin{equation}\label{eq/NonUniform/33}
s(\vR) \equiv -\textstyle{\frac{\partial}{\partial T}}\,f(T, \, v(\vR), \, \arglistr{\xi}{n-1}) = - \textstyle{\frac{\partial}{\partial T}}\, f_{0}(T, \, v(\vR), \, \arglistr{\xi}{n-1})
\end{equation}%
\eqr{eq/NonUniform/34a} has the form of the usual Gibbs relation for a homogeneous mixture. For an inhomogeneous mixture, however, the validity of such a relation is not obvious.
\eqr{eq/NonUniform/34a} implies that with respect to the variations of the thermodynamic variables the specific Helmholtz energy is still homogeneous of the first order. We will call
\eqr{eq/NonUniform/34a} the ordinary Gibbs relation. With the help of \eqr{eq/NonUniform/15} and \eqr{eq/NonUniform/34a} we obtain the Gibbs-Duhem relation
\begin{equation}\label{eq/NonUniform/34gd}
s(\vR)\,\delta T - v(\vR)\,\delta p\,(\vR)  + \delta\mu_{m} + {\sum\limits_{i=1}^{n-1}{\xi _{i}(\vR)\,\delta\psi_{i}}} = 0 %
\end{equation}%

Using the following conditions, which are true for equilibrium
\begin{equation}\label{eq/NonUniform/36}
\begin{array}{rl}
\nabla\,T(\vR) &= 0 \\
\nabla\,\mu_{n}(\vR) &= 0 \\
\nabla\,\psi_{i}(\vR) &= 0, \quad \text{for } i=\nmorange \\
\nabla_{\alpha}\,\sigma_{\alpha\beta}(\vR)  &= 0 %
\end{array}
\end{equation}%
and \eqr{eq/NonUniform/15} together with \eqr{eq/NonUniform/11} we obtain
\begin{equation}\label{eq/NonUniform/35a}
\nabla_{\beta}\,f(\vR) = - p\,(\vR)\,\nabla_{\beta}\,v(\vR) + {\sum\limits_{i=1}^{n-1}{\psi_{i}\,\nabla_{\beta}\,\xi_{i}(\vR)}} + v(\vR)\,\nabla_{\alpha}\,\gamma_{\alpha \beta}(\vR) %
\end{equation}%
We will call \eqr{eq/NonUniform/35a} the spatial Gibbs relation. As the temperature is independent of the position, the expected $-s(\vR)\,\nabla_{\beta}T$ term is zero.

\subsection{Equilibrium surface.}\label{sec/NonUniform/Matter/Surface}

Away from the surface $c(\vR)\rightarrow c$, $\xi _{k}(\vR)\rightarrow \xi _{k}$ and $\nabla{c(\vR)} \rightarrow 0$, $\nabla{\xi_{k}(\vR)} \rightarrow 0$ and we have the homogeneous mixture.
Thus, the usual thermodynamic relations for the homogeneous mixture are valid. The specific Helmholtz energy is given by
\begin{equation}\label{eq/Uniform/07a}
f_{0}(\alTcxi) = \mu_{n,\,0}(\alTcxi) + {\sum\limits_{k=1}^{n-1}{\psi_{k,\,0}(\alTcxi)\,\xi_{k}}} - p_{0}\,(\alTcxi)\,v
\end{equation}
with the specific entropy, pressure and chemical potentials given by
\begin{equation}\label{eq/Uniform/06}
\begin{array}{rl}
s_{0}(\alTcxi) &= -{\frac{\partial }{\partial T}}\,f_{0}(\alTcxi) \\
\\
p_{0}\,(\alTcxi) &= -{\frac{\partial }{\partial v}}\,f_{0}(\alTcxi) \\
\\
\psi_{k,\,0}(\alTcxi) &= {\frac{\partial }{\partial \xi_{k}}}\,f_{0}(\alTcxi),\;k=\nmorange%
\end{array}
\end{equation}
such that the variation of the specific Helmholtz energy
\begin{equation}\label{eq/Uniform/05}
df_{0}(\alTcxi) = -s_{0}(\alTcxi)\,dT - p_{0}(\alTcxi)\,dv + {\sum\limits_{k=1}^{n-1}{\psi_{k,\,0}(\alTcxi)\,d\xi_{k}}}
\end{equation}
where $\psi_{k,\,0}(\alTcxi) \equiv \mu_{k,\,0}(\alTcxi) - \mu_{n,\,0}(\alTcxi)$.

Quantities derived in \ssecr{sec/NonUniform/Matter/Lagrange} are converged in a homogeneous limit in a following way
\begin{equation}\label{eq/NonUniform/14}
\begin{array}{rllll}
\SumVarea \psi_{k} & \rightarrow & \psi_{k,\,0}(\alTcxi) & = & {\frac{\partial}{\partial \xi_{k}}}f_{0}(\alTcxi) \\
\SumVarea \mu_{n} & \rightarrow & \mu_{n,\,0}(\alTcxi) & = & f_{0}(\alTcxi) - {\sum\limits_{i=1}^{n-1}{\psi_{i,\,0}(\alTcxi)\xi_{i}}} + p_{0}(\alTcxi)\,v \\
\SumVarea p\,(\vR) & \rightarrow & p_{0}(\alTcxi) & = & c^{2}{\frac{\partial}{\partial c}}f_{0}(\alTcxi) \\
\SumVarea \sigma_{\alpha\beta}(\vR) & \rightarrow & \sigma_{\alpha\beta,\,0}(\alTcxi) & = & p_{0}(\alTcxi)\,\delta_{\alpha\beta}%
\end{array}
\end{equation}%
We use these limits to determine the meaning of the derived quantities in the interfacial region, where gradients are not negligible.

In equilibrium $\psi_{k}$ and $\mu_{n}$ are everywhere constant. Away from the surface they represent the homogeneous chemical potentials, which in equilibrium should be constant everywhere,
particulary throughout the whole interfacial region. Thus it is natural to identify $\psi_{k}$ and $\mu_{n}$ with the chemical potentials also within the interfacial region.

Before determine the meaning of $p(\vR)$ and $\sigma_{\alpha\beta}(\vR)$ we have to resolve an ambiguity in the definition of $\sigma_{\alpha\beta}(\vR)$. It follows from \eqr{eq/NonUniform/10}
that a constant tensor can be added to $\sigma_{\alpha\beta}(\vR)$ without affecting the validity of this equation. In a homogeneous limit this tensor does not vanish, so it should be present in
the homogeneous tensor $\sigma_{\alpha\beta,\,0}(\vR)$, which is proportional to the homogeneous pressure $p_{0}(\alTcxi)$. The homogeneous pressure $p_{0}(\alTcxi)$, however, is determined
unambiguously by the specified equations of state. It follows then that this constant tensor have to be equal to zero and $\sigma_{\alpha\beta}(\vR)$ is given by \eqr{eq/NonUniform/11}
unambiguously.

Using that $\psi_{k}$ and $\mu_{n}$ are the chemical potentials and \eqr{eq/NonUniform/15} it is then also natural to identify $p(\vR)$ given by \eqr{eq/NonUniform/09} with a pressure
everywhere. This pressure is not constant, however. The tensor $\sigma_{\alpha\beta}(\vR)$ can be identified with the tensorial pressure. It is known that at the surface one can speak about the
"parallel" and the "perpendicular" pressure \cite{RowlinsonWidom}, so the pressure reveals tensorial behavior. For a flat surface, when all the properties change in one direction, say $x$, one
can show that $\sigma_{xx}(\vR)$ is the "perpendicular" pressure and $p(\vR) = \sigma_{yy}(\vR) = \sigma_{zz}(\vR)$ is the "parallel" pressure. For curved surfaces such an identification,
however, can in general not be made.

One can also conclude, that the quantity, determined by \eqr{eq/NonUniform/33} is the specific entropy of the mixture in the interfacial region. It does not have gradient contributions. This is
due to the assumption that the coefficients of the square gradient contributions are independent of temperature. We refer to van der Waals \cite{vdW/sg, vdW/translation, RowlinsonWidom} who
discussed this property.

We shall also define other thermodynamic potentials in the square gradient model for the interfacial region. Considering \eqr{eq/NonUniform/15} and conforming to \eqr{eq/NonUniform/14} we define
interfacial molar internal energy, enthalpy and Gibbs energy densities as follows
\begin{equation}\label{eq/NonUniform/41}
\begin{array}{l}
\BigVarea u(\vR) = f(\vR) + s(\vR)\,T\\
\BigVarea h(\vR) = u(\vR) + p\,(\vR)\,v(\vR)\\
\BigVarea g(\vR) = f(\vR) + p\,(\vR)\,v(\vR) %
\end{array}
\end{equation}
It is important to realize that these thermodynamic relations are true in the interfacial region only by definition. One can also find support for these definitions in \cite{bedeaux/vdW/I} where
they were considered for a simplified one-component system.

Using \eqr{eq/NonUniform/34a} and \eqr{eq/NonUniform/41}, for the internal energy at each point in space we then get the ordinary Gibbs relation
\begin{equation}\label{eq/NonUniform/34}
\delta u(s(\vR), v(\vR), \arglistr{\xi}{n-1}) = T\,\delta s(\vR) - p\,(\vR)\,\delta v(\vR) + {\sum\limits_{i=1}^{n-1}{\psi_{i}\,\delta \xi_{i}(\vR)}}
\end{equation}%
From \eqr{eq/NonUniform/36} and \eqr{eq/NonUniform/41} we get the spatial Gibbs relation
\begin{equation}\label{eq/NonUniform/35}
\nabla_{\beta}\,u(\vR) = T\,\nabla_{\beta}\,s(\vR) - p\,(\vR)\,\nabla_{\beta}\,v(\vR) + {\sum\limits_{i=1}^{n-1}{\psi_{i}\,\nabla_{\beta}\,\xi_{i}(\vR)}} + v(\vR)\,\nabla_{\alpha}\,\gamma_{\alpha
\beta}(\vR)
\end{equation}%
One can easily write the Gibbs relations for other thermodynamic potentials.

\section{Gradient model for the specific variables per unit of volume.}\label{sec/NonUniform/Volume}

We write the extended Helmholtz energy per unit of volume as
\begin{equation}\label{eq/NonUniform/19}
f^{v}(\vR) = f_{0}^{v}(T,\,c) + \Kappa^{v}(c,\,\nabla{c})
\end{equation}%
where
\begin{equation}\label{eq/NonUniform/26}
\Kappa^{v}(c,\,\nabla{c}) \equiv {\frac{1}{2}}{\sum\limits_{i,j=1}^{n}{\kappa_{ij}^{v}\,\nabla{c_{i}(\vR)}\spd\nabla{c_{j}(\vR)}}}%
\end{equation}
We use $c$ as short notation instead of whole set $\{\arglist{c}{n}\}$ and $\nabla{c}$ as short notation instead of $\{\arglist{\nabla{c}}{n}\}$ in \secr{sec/NonUniform/Volume}. It should not be
confused with the total molar concentration, used in \secr{sec/NonUniform/Matter}.

The coefficients $\kappa_{ij}^{v}$ are different from those we used in \secr{sec/NonUniform/Matter}. One can derive the following relations between them.
\begin{equation}\label{}
\begin{array}{rl}
\SumVarea \kappa &= {\sum\limits_{i,j=1}^{n-1}{\xi_{i}\,\xi_{j}\,\mathrm{k}_{ij}^{v}}} + 2{\sum\limits_{i}^{n-1}{\xi_{i}\,\mathrm{k}_{i}^{v}}} + \mathrm{k}^{v}
\\
\SumVarea \kappa_{i} &= c\,{\sum\limits_{j=1}^{n-1}{\xi_{j}\,\mathrm{k}_{ij}^{v}}} + c\,\mathrm{k}_{i}^{v}
\\
\SumVarea \kappa_{ij} &= c^{2}\,\mathrm{k}_{ij}^{v}
\end{array}
\end{equation}
where
\begin{equation}\label{}
\begin{array}{rl}
\BigVarea \mathrm{k}_{ij}^{v} &\equiv \kappa_{ij}^{v}+\kappa_{nn}^{v}-\kappa_{in}^{v}-\kappa_{nj}^{v}
\\
\BigVarea \mathrm{k}_{i}^{v} &\equiv \kappa_{in}^{v}-\kappa_{nn}^{v}
\\
\BigVarea \mathrm{k}^{v} &\equiv \kappa_{nn}^{v}
\end{array}
\end{equation}
and
\begin{equation}\label{}
\SumVarea \kappa_{ij}^{v} = \kappa + {\frac{1}{{\mathrm{c}}}}(\kappa_{i}+\kappa_{j}) +
{\frac{1}{{\mathrm{c}}^{2}}}\Big(\kappa_{ij}-2\sum\limits_{\ell=1}^{n}{\kappa_{\ell}c_{\ell}}\Big) - %
{\frac{1}{{\mathrm{c}}^{3}}}\sum\limits_{\ell=1}^{n}{(\kappa_{\ell j}+\kappa_{i \ell})c_{\ell}} +
{\frac{1}{{\mathrm{c}}^{4}}}\sum\limits_{\ell_{1},\ell_{2}=1}^{n}{\kappa_{\ell_{1}\ell_{2}}c_{\ell_{1}}c_{\ell_{2}}}
\end{equation}
where $\mathrm{c}\equiv {\sum\limits_{\ell=1}^{n}{c_{\ell}}}$.

We will not repeat in details the procedure, given in \secr{sec/NonUniform/Matter} and will give only the results here. Using the Lagrange method we obtain the expressions for the constant
Lagrange multipliers, which are equal to the chemical potentials
\begin{equation}\label{eq/NonUniform/24}
\mu_{k} = {\frac{\partial}{\partial c_{k}}}(f_{0}^{v}+\Kappa^{v}) - {\sum\limits_{i=1}^{n}{\nabla\spd(\kappa_{ik}^{v}\,\nabla{c_{i}})}}
\end{equation}%
and the expression for a pressure $p$:
\begin{equation}\label{eq/NonUniform/25}
p(\vR) = {\sum\limits_{i=1}^{n}{c_{i}\,{\frac{\partial}{\partial c_{i}}}(f_{0}^{v} + \Kappa^{v}) }} - \big(f_{0}^{v} + \Kappa^{v}\big) -
{\sum\limits_{i,j=1}^{n}{c_{j}\,\nabla\spd(\kappa_{ij}^{v}\,\nabla{c_{i}})}}
\end{equation}%
which was defined by
\begin{equation}\label{eq/NonUniform/32}
f^{v}(\vR)={\sum\limits_{i=1}^{n}{\mu _{i}\,c_{i}(\vR)}}-p\,(\vR)
\end{equation}%
One can derive the same symmetric tensorial pressure $\sigma_{\alpha\beta}(\vR)$ as in \eqr{eq/NonUniform/11}, which obeys the relation \eqr{eq/NonUniform/10}, where the symmetric tension tensor
$\gamma_{\alpha\beta}(\vR)$ is given by
\begin{equation}\label{eq/NonUniform/29}
\gamma_{\alpha\beta}(\vR) = {\sum\limits_{i,j=1}^{n}{\kappa_{ij}^{v}\,{\PCDF{c_{i}(\vR)}{\alpha}}{\PCDF{c_{j}(\vR)}{\beta}}}}
\end{equation}%

Varying the total Helmholtz energy $\delta F[T,\,c(\vR),\,\nabla{c(\vR)}]$ with respect to the variation of the variables we obtain the total thermodynamic differential of the specific Helmholtz
energy as
\begin{equation}\label{eq/NonUniform/61}
\delta f^{v}(T,\,c,\,\nabla{c}) = {\frac{\partial f_{0}^{v}}{\partial T}}\,\delta T + {\sum\limits_{k=1}^{n}{\mu_{k}\,\delta c_{k}}} + \nabla\spd{\delta\mathbf\Theta}^{v}
\end{equation}
where
\begin{equation}\label{eq/NonUniform/63}
{\delta\mathbf\Theta}^{v}(c,\,\nabla{c}) \equiv {\sum\limits_{i,j=1}^{n}{(\kappa_{ij}\,\nabla{c_{i}})\,\delta c_{j}}}
\end{equation}
The total Helmholtz energy variation becomes then
\begin{equation}\label{eq/NonUniform/64}
\delta F[T,\,c,\,\nabla{c}] = \int_{V}d\vR\,\Big({\frac{\partial f_{0}^{v}}{\partial T}}\,\delta T + {\sum\limits_{k=1}^{n}{\mu_{k}\,\delta c_{k}}}\Big)
\end{equation}
since the boundary integral $\int_{S}dS\,\vn_{s}\spd{\delta\mathbf\Theta}^{v}$ disappears. Thus, we will interpret the expression in parenthesis as the total thermodynamic differential of the
specific Helmholtz energy:
\begin{equation}\label{eq/NonUniform/66}
\delta f^{v}(T,\,c,\,\nabla{c}) = {\frac{\partial f_{0}^{v}}{\partial T}}\,\delta T + {\sum\limits_{k=1}^{n}{\mu_{k}\,\delta c_{k}}}
\end{equation}

We write \eqr{eq/NonUniform/66} in a form
\begin{equation}\label{eq/NonUniform/43a}
\delta f^{v}(T, \arglistr{c}{n}) = -s^{v}(\vR)\,\delta T + {\sum\limits_{i=1}^{n}{\mu_{i}\,\delta c_{i}(\vR)}} %
\end{equation}%
which we will call \eqr{eq/NonUniform/43a} the ordinary Gibbs relation. Here
\begin{equation}\label{eq/NonUniform/44}
s^{v}(\vR) \equiv - \textstyle{\frac{\partial}{\partial T}}\, f^{v}(T,\,\arglistr{c}{n}) = - \textstyle{\frac{\partial}{\partial T}}\, f_{0}^{v}(T,\,\arglistr{c}{n})
\end{equation}%
is the specific entropy of the mixture.  With the help of \eqr{eq/NonUniform/32} and \eqr{eq/NonUniform/43a} we obtain the Gibbs-Duhem relation
\begin{equation}\label{eq/NonUniform/43gd}
s^{v}(\vR)\,\delta T - \delta p\,(\vR)  + {\sum\limits_{k=1}^{n}{c_{k}(\vR)\,\delta\mu_{k}}} = 0 %
\end{equation}%

Using the following conditions, which are true for equilibrium
\begin{equation}\label{eq/NonUniform/36a}
\begin{array}{rl}
\nabla\,T(\vR) &= 0 \\
\nabla\,\mu_{k}(\vR) &= 0, \quad \text{for } k=\nrange \\
\nabla_{\alpha}\,\sigma_{\alpha\beta}(\vR)  &= 0 %
\end{array}
\end{equation}%
and \eqr{eq/NonUniform/32} together with \eqr{eq/NonUniform/11} we obtain the spatial Gibbs relation
\begin{equation}\label{eq/NonUniform/45a}
\nabla_{\beta}\,f^{v}(\vR) = {\sum\limits_{i=1}^{n}{\mu_{i}\,\nabla_{\beta}\,c_{i}(\vR)}} + \nabla_{\alpha}\,\gamma_{\alpha\beta}(\vR) %
\end{equation}%

The interfacial specific internal energy, enthalpy and Gibbs energy densities are
\begin{equation}\label{eq/NonUniform/42}
\begin{array}{l}
\SumVarea u^{v}(\vR) = f^{v}(\vR) + s^{v}(\vR)\,T  \\
\SumVarea h^{v}(\vR) = s^{v}(\vR)\,T + {\sum\limits_{i=1}^{n}{\mu_{i}\,c_{i}(\vR)}} \\
\SumVarea g^{v}(\vR) = {\sum\limits_{i=1}^{n}{\mu_{i}\,c_{i}(\vR)}} %
\end{array}
\end{equation}

The Gibbs relations for the internal energy are
\begin{equation}\label{eq/NonUniform/43}
\delta u^{v}(s(\vR), \arglistr{c}{n}) = T\,\delta s^{v}(\vR) + {\sum\limits_{i=1}^{n}{\mu_{i}\,\delta c_{i}(\vR)}}
\end{equation}%
and
\begin{equation}\label{eq/NonUniform/45}
\nabla_{\beta}\,u^{v}(\vR) = T\,\nabla_{\beta}\,s^{v}(\vR) + {\sum\limits_{i=1}^{n}{\mu_{i}\,\nabla_{\beta}\,c_{i}(\vR)}} + \nabla_{\alpha}\,\gamma_{\alpha \beta}(\vR)
\end{equation}%

\section{Typical profiles for the binary mixture.}\label{sec/Profiles}

\begin{figure}[hbt!]
\centering
\includegraphics[scale=0.7]{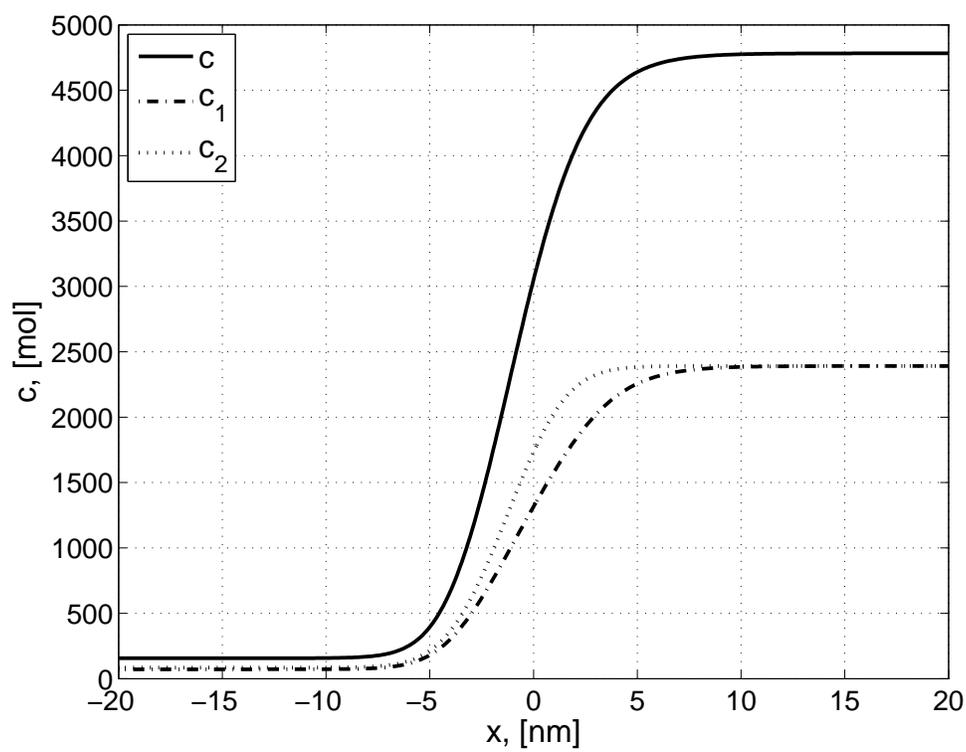}
\caption{Molar concentration profile}\label{Fig_c}
\end{figure}

\begin{figure}[hbt!]
\centering
\includegraphics[scale=0.7]{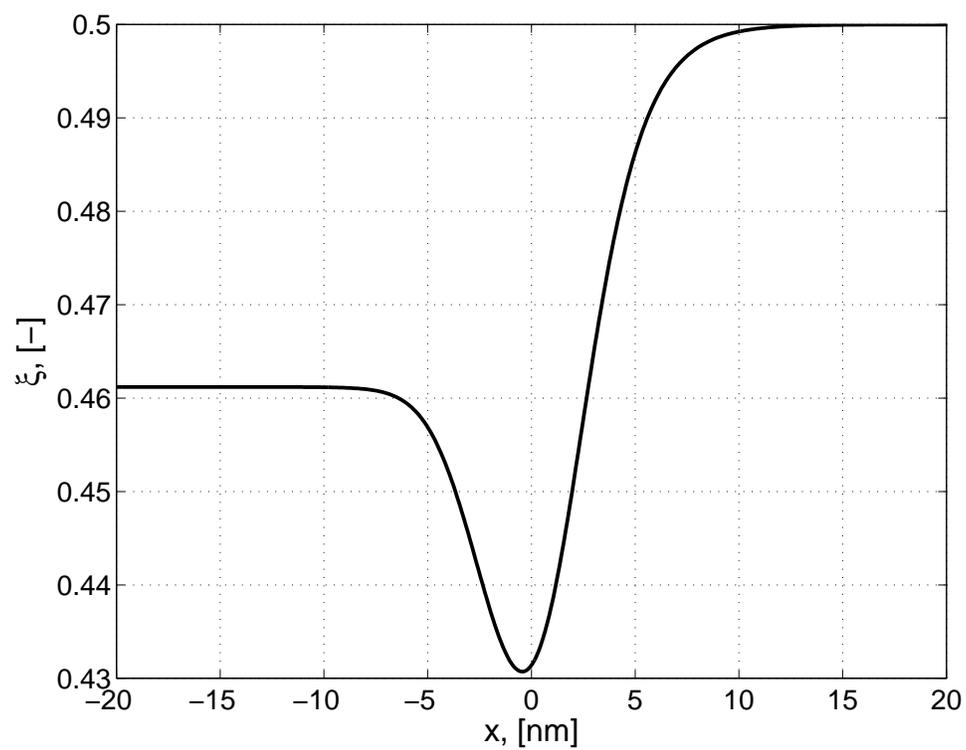}
\caption{Molar fraction profile}\label{Fig_xi}
\end{figure}

\begin{figure}[hbt!]
\centering
\includegraphics[scale=0.7]{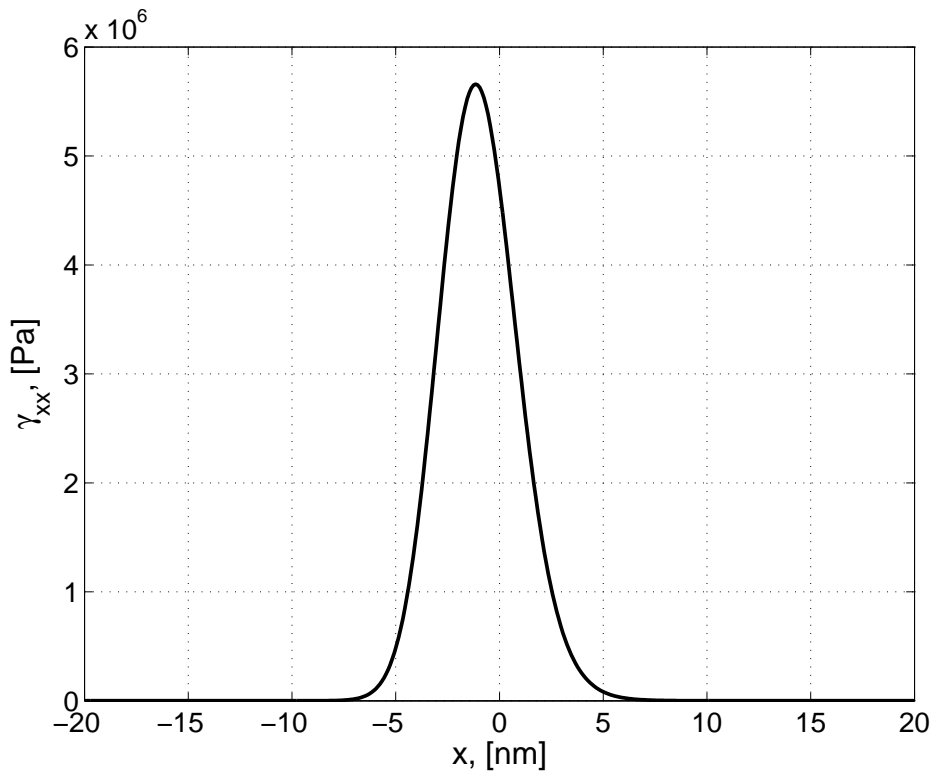}
\caption{Tension $\gamma_{xx}$ profile}\label{Fig_gamma}
\end{figure}

In order to illustrate the results, which one can obtain using the above procedure, we have applied it to a special case. This requires a number of approximations, connected with the specific
mixture and the geometry. We consider a flat liquid-vapor interface of the binary mixture of cyclohexane (1\tsup{st} component) and $n$-hexane (2\tsup{nd} component). The equilibrium profiles
are obtained using \eqr{eq/NonUniform/08}. The values for the coexisting chemical potentials are obtained from the coexisting conditions using the van-der-Waals equation of state. For the binary
mixture this requires two independent thermodynamic properties, which we have chosen to be the equilibrium temperature $T = 330$ K and molar fraction of the first component in the liquid phase
$\xi_l = 0.5$. As cyclohexane and $n$-hexane are rather similar, the gradient coefficients $\kappa_{11}^{v}$, $\kappa_{22}^{v}$ and $\kappa_{12}^{v}$ were chosen to be equal. We use the value
12E-18 J*m\tsup{5}/mol\tsup{2} which reproduces the experimental value of the surface tension 0.027 N/m. The resulting profiles for the molar concentration, the mole fraction of the first
component and the tension tensor component $\gamma_{xx}$, the integral of which gives the surface tension, are given in Figs.~\ref{Fig_c}-\ref{Fig_gamma}. This paper is the first in a sequence
of papers. In the second paper we will extend the present analysis to non-equilibrium mixtures. Details of the numerical procedure will be given in a third paper which discusses the validity of
local equilibrium for the Gibbs surface.


\section{Discussion and conclusions.}\label{sec/Discussion}

In this paper we have established the framework of the gradient model for the liquid-vapor (or, alternatively, liquid-liquid) interface in an isotropic non-polarizable mixture. It is necessary
that the homogeneous Helmholtz energy $f_{0}$ allows solutions which imply equilibrium coexistence between different phases. Otherwise we only have the homogeneous phase. Standard mixture
theories \cite{sengers/EOS} give a Helmholtz energy which allows liquid-vapor coexistence.

Using the assumption that in the interfacial region the fluid can be described by the local densities and their gradients, we have extended the gradient models, used to describe one-component
fluids and binary mixtures, to three-dimensional multi-component mixtures. The condition which the system should satisfy is that the total Helmholtz energy is minimal. With the help of Lagrange
method is was possible to derive the equations, which the profile distribution should satisfy, given the fixed total molar content of the components. The Lagrange multipliers are equal to the
chemical potentials of the coexisting liquid and vapor. It was also possible to determine the pressure behavior in the interfacial region. It is crucial, that the pressure has a tensorial
behavior. The difference between the tensorial part of the pressure tensor and the scalar part determines the surface tension.

An important observation is the ambiguity in the determination of the local thermodynamic potentials, for instance the specific Helmholtz energy. While the total Helmholtz energy is unique and
has a minimum, the specific Helmholtz energy is not unique. One can add a term which is the divergence of some vector field, if the normal component of this field vanishes on the system
boundaries, without affecting the total Helmholtz energy. This general observation in the context of the gradient model implies that the density gradients are taken equal to zero on the
boundary. It must be emphasized, that for realistic boundaries these gradients are not zero. We refer to the wall-theorem \cite{lebowitz/wall} in this context. We take the freedom to set them
equal to zero assuming, that the boundary layer does not affect the properties of the interface we want to study. In the \appr{sec/Ambiguity} we show that this ambiguity does not affect the
results.

As one can see from \eqr{eq/NonUniform/15} and \eqr{eq/NonUniform/32}, it is possible to relate the thermodynamic variables for an inhomogeneous fluid in the same way as it is done for a
homogeneous one. However, unlike the homogeneous mixture, these variables contain gradient contributions. This means that local equilibrium for such a system is not satisfied. The local behavior
of the mixture is determined not only by it's local properties but also by it's nearest surroundings. Moreover, in contrast to the homogeneous description, the local properties now vary in the
space.

We have given explicit expressions for each thermodynamic quantity in the interfacial region. We have also determined how the thermodynamic potentials change with the change of the variables
they depend on. An important part of the thermodynamic description is the relations between the rate of change of the thermodynamic variables, the Gibbs relations. In contrast to a homogeneous
system, for an inhomogeneous system, in particular the interfacial region, thermodynamic variables vary also in space. Thus one can speak about the relation between the rates of change of
thermodynamic variables for a given point in space, the ordinary Gibbs relation. One has also to speak about the rates of change of the thermodynamic variables in space, the spacial Gibbs
relation. Even though the thermodynamic potentials, particularly the specific Helmholtz energy, depend on the spatial derivatives of the densities, we have shown that variation of these
gradients do not contribute to the ordinary Gibbs relations. Thus, the ordinary Gibbs relations have the ordinary form of the Gibbs relations for the homogeneous mixture. The important
observation here is, however, that the ordinary Gibbs relations relate the inhomogeneous thermodynamic variables, i.e. those, which contain the gradient contribution. As the spatial derivatives
of the temperature, chemical potentials and pressure tensor are zero in equilibrium, we can determine the spatial Gibbs relation. The new term which appears because of the inhomogeneity is
$\nabla_{\alpha}\,\gamma_{\alpha\beta}(\vR)$, which is only unequal to zero close to the surface.

For temperatures far from the critical point the surface thickness is known to be very small (in the sub-nm range). This imposes an upper limit to the values of the coefficients $\kappa$,
$\kappa_{i}$, $\kappa_{ij}$ and $\kappa_{ij}^{v}$.

In previous sections we used different specific variables. For molar densities we had the coefficients $\kappa$, $\kappa_{i}$ and $\kappa_{ij}$ and for densities per unit of volume
$\kappa_{ij}^{v}$. One can determine the relations between these coefficients and verify, that all the quantities, determined in \secr{sec/NonUniform/Matter} and \secr{sec/NonUniform/Volume} are
the same. Thus, \eqr{eq/NonUniform/09} and \eqr{eq/NonUniform/25} give the same quantity $p\,(\vR)$, \eqr{eq/NonUniform/12} and \eqr{eq/NonUniform/29} -- the same $\gamma_{\alpha\beta}(\vR)$. And
$\sigma_{\alpha\beta}(\vR)$, which is given by \eqr{eq/NonUniform/11} is the same for both sets of variables. $\mu_{n}$ in \eqr{eq/NonUniform/08} and \eqr{eq/NonUniform/24} is the same and
$\psi_{k}$ taken from \eqr{eq/NonUniform/08} are equal to $\mu_{k}-\mu_{n}$ taken from \eqr{eq/NonUniform/24}. This shows that the inhomogeneous equilibrium description is independent of the
choice of independent variables. This is similar to the description of the homogeneous equilibrium phase.

The analysis in this paper gives the basis to extend the description to non-equilibrium systems. For one-component systems, in which the properties varied only in one direction, such an extension
was given by Bedeaux et. al. \cite{bedeaux/vdW/I, bedeaux/vdW/II, bedeaux/vdW/III}.

\begin{acknowledgments}
One of the authors (D.B.) wants to thank Edgar Blokhuis for help and advise.
\end{acknowledgments}

\appendix

\section{On the ambiguity in the specific quantities.}\label{sec/Ambiguity}

We show here that ambiguities present in the definition of the specific Helmholtz energy and the total thermodynamic differential of the specific Helmholtz energy do not affect the validity of
all the thermodynamic relations, derived in the present article. We do this for molar specific variables, the same arguments can be used for the specific variables per unit of volume.

The total Helmholtz energy of a mixture $\int_{V}d\vR\,c(\vR)f(\vR)$ can be expanded around homogeneous state as
\begin{equation}\label{}
\begin{array}{ll}
F = \int_{V}d\vR\,\Big[&c\,f_{0}(T,\,c,\,\xi) + \\
&+ {\frac{1}{2}}\kappa^{(1)}\,|\nabla{c}\,|^{2} + {\sum\limits_{i=1}^{n-1}{\kappa_{i}^{(1)}\,\nabla{c}\spd\nabla{\xi_{i}}}} +
{\frac{1}{2}}{\sum\limits_{i,j=1}^{n-1}{\kappa_{ij}^{(1)}\,\nabla{\xi_{i}}\spd\nabla{\xi_{j}}}} + \\
&+ \kappa^{(2)}\,\nabla^{2}{c} + {\sum\limits_{i=1}^{n-1}{\kappa_{i}^{(2)}\,\nabla^{2}{\xi_{i}}}} + \ldots\Big]
\end{array}
\end{equation}%
where the series is truncated after the second order terms. As it is shown in \eqr{eq/NonUniform/01a} this can be rearranged as following:
\begin{equation}\label{}
\begin{array}{ll}
F = \int_{V}d\vR\,\Big[&c\,f_{0}(T,\,c,\,\xi) + \\
&+ {\frac{1}{2}}\kappa\,|\nabla{c}\,|^{2} + {\sum\limits_{i=1}^{n-1}{\kappa_{i}\,\nabla{c}\spd\nabla{\xi_{i}}}} + {\frac{1}{2}}{\sum\limits_{i,j=1}^{n-1}{\kappa_{ij}\,\nabla{\xi_{i}}\spd\nabla{\xi_{j}}}}  + \\
&+ \nabla\spd\big(\kappa^{(2)}\nabla{c} + {\sum\limits_{i=1}^{n-1}{\kappa_{i}^{(2)}\,\nabla{\xi_{i}}}}\big) + \ldots\Big]
\end{array}
\end{equation}%
where
\begin{equation}\label{}
\begin{array}{ll}
\kappa &= \kappa^{(1)}-2\,{\frac{\partial \kappa^{(2)}}{\partial c}} \\
\kappa_{i} &= \kappa_{i}^{(1)} - {\frac{\partial \kappa^{(2)}}{\partial \xi_{i}}} - {\frac{\partial \kappa^{(2)}_{i}}{\partial c}} \\
\kappa_{ij} &= \kappa_{ij}^{(1)} - {\frac{\partial \kappa^{(2)}_{i}}{\partial \xi_{j}}}- {\frac{\partial \kappa^{(2)}_{j}}{\partial \xi_{i}}}
\end{array}
\end{equation}%
are the coefficients used in \eqr{eq/NonUniform/08a}.

Let us define
\begin{equation}\label{}
\begin{array}{l}
f_{1}(\vR) \equiv c\,f_{0}(T,\,c,\,\xi) + {\frac{1}{2}}\kappa\,|\nabla{c}\,|^{2} + {\sum\limits_{i=1}^{n-1}{\kappa_{i}\,\nabla{c}\spd\nabla{\xi_{i}}}} +
{\frac{1}{2}}{\sum\limits_{i,j=1}^{n-1}{\kappa_{ij}\,\nabla{\xi_{i}}\spd\nabla{\xi_{j}}}} \\
f_{2}(\vR) \equiv f_{1}(\vR)  + \nabla\spd\big(\kappa^{(2)}\nabla{c} + {\sum\limits_{i=1}^{n-1}{\kappa_{i}^{(2)}\,\nabla{\xi_{i}}}}\big)\\
\end{array}
\end{equation}%
so that $F = \int_{V}d\vR\,c(\vR)f_{1}(\vR) = \int_{V}d\vR\,c(\vR)f_{2}(\vR)$, since $\int_{V}d\vR\,c(\vR)\big(f_{2}(\vR)-f_{1}(\vR)\big) = 0$. One can also define
\begin{equation}\label{}
\begin{array}{ll}
p_{1}(\vR) &\equiv \mu_{n} + {\sum\limits_{i=1}^{n-1}{\psi_{i}\,\xi_{i}(\vR)}} - f_{1}(\vR) = c^{2}{\frac{\partial}{\partial c}}\big(f_{0} + \Kappa\big) - c\,\nabla\spd\Big(\kappa\,\nabla{c} + {\sum\limits_{i=1}^{n-1}{\kappa_{i}\,\nabla{\xi_{i}}}}\Big) \\
p_{2}(\vR) &\equiv \mu_{n} + {\sum\limits_{i=1}^{n-1}{\psi_{i}\,\xi_{i}(\vR)}} - f_{2}(\vR) = p_{1}(\vR) - \nabla\spd\big(\kappa^{(2)}\nabla{c} + {\sum\limits_{i=1}^{n-1}{\kappa_{i}^{(2)}\,\nabla{\xi_{i}}}}\big) %
\end{array}
\end{equation}%
and follow the same procedure as in \ssecr{sec/NonUniform/Matter/Lagrange}. (Note, that the second order terms should be added to the Euler-Lagrange equations in thus case). Then, as it follows
from the known theorem of the variational calculus, one obtains for the chemical potentials $\mu_{n}$ and $\psi_{k}$
\begin{equation}\label{}
\begin{aligned}
\mu_{n} & =  {\frac{\partial}{\partial c}}\Big(c\,f_{1}\Big) - {\sum\limits_{i=1}^{n-1}{\psi_{i}\xi _{i}}} - \nabla\spd\Big(\kappa\,\nabla{c} + {\sum\limits_{i=1}^{n-1}{\kappa_{i}\,\nabla{\xi_{i}}}} \Big) \\%
\psi_{k} & = {\frac{\partial}{\partial \xi_{k}}}\Big(f_{1}\Big) - {\frac{1}{c}}\,\nabla\spd\Big(\kappa_{k}\,\nabla{c} + {\sum\limits_{i=1}^{n-1}{\kappa_{ik}\,\nabla{\xi_{i}}}}\Big) , \quad k=\nmorange %
\end{aligned}
\end{equation}%
and for the pressure tensor $\sigma_{\alpha\beta}(\vR)$
\begin{equation}\label{}
\sigma_{\alpha\beta}(\vR) = p_{1}(\vR)\,\delta_{\alpha\beta} + \gamma_{\alpha\beta}(\vR)
\end{equation}%
for the both definitions $p_{1}(\vR)$ and $p_{2}(\vR)$. As one can see, there is no ambiguity in the expressions for the chemical potentials and the pressure tensor. They are determined
irrespectively of the ambiguity in the definition of the specific Helmholtz energy. Thus, it is natural to use $f_{1}(\vR)$ as the specific Helmholtz energy and use $p_{1}(\vR)$ as a pressure.
As we could see in \ssecr{sec/NonUniform/Matter/Surface} only $p_{1}(\vR)$ has a physical meaning but not $p_{2}(\vR)$. With such a choice all the thermodynamic quantities, derived in the paper
do not contain any ambiguity.

\section{Symbols list and notation convention.}\label{sec/Symbols}

\begin{longtable}{c@{\quad--\quad}p{.8\linewidth}}
$\cdot$ & contraction sign\\
$\nrange$ & enumeration of all integers from $1$ to $n$\\
$\alpha, \beta$ & cartesian indices \\
$\delta$ & total thermodynamic differential \\
$\delta_{\alpha\beta}$ & Kroneker symbol\\
\end{longtable}

\begin{longtable}{l@{\quad}l@{\quad--\quad}p{.8\linewidth}}
$\nabla$ & [ 1/m ] & nabla operator\\
$\nabla_{\alpha}$ & [ 1/m ] & partial derivative with respect to $x_{\alpha}$\\
$\gamma_{\alpha\beta}$ & [ Pa ] & tension tensor \\
$\Delta$ & [ 1/m\tsup{2} ] & Laplace operator\\
$\Kappa$ & [ J/mol ] & square gradient contribution\\
$\Kappa^{v}$ & [ J/m\tsup{3} ] & -------- $>>$ -------- $>>$ --------\\
$\kappa$ & [ J m\tsup{5}/mol\tsup{2} ] & square gradient coefficients for molar units\\
$\kappa_{i}$ & [ J m/mol ] & -------- $>>$ -------- $>>$ --------\\
$\kappa_{ij}$ & [ J/m\tsup{3} ] & -------- $>>$ -------- $>>$ --------\\
$\kappa_{ij}^{v}$ & [ J m\tsup{5}/mol\tsup{2} ] & square gradient coefficients for volume units\\
$\mu_{k}$ & [ J/mol ] & molar chemical potential of component $k$\\
$\mu_{k}^{\measure}$ & [ J/kg ], [ J/mol ] & matter chemical potential of component $k$\\
$\mu_{k,\,0}$ & [ J/mol ] & homogeneous molar chemical potential of component $k$\\
$\mu_{k,\,0}^{\measure}$ & [ J/kg ], [ J/mol ] & homogeneous matter chemical potential of component $k$\\
$\nu$ & [ mol ] & total number of moles \\
$\nu_{k}$ & [ mol ] & number of moles of component $k$\\
$\rho$ & [ kg/m\tsup{3} ] & total mass density\\
$\rho_{k}$ & [ kg/m\tsup{3} ] & mass density of component $k$\\
$\sigma_{\alpha\beta}$ & [ Pa ] & pressure tensor\\
$\xi$ & [ -- ] & set of molar fractions $\{\arglist{\xi}{n-1}\}$\\
$\xi_{k}$ & [ -- ] & molar fraction of component $k$\\
$\xi_{k}^{\measure}$ & [ -- ] & matter fraction of component $k$\\
$\psi_{k}$ & [ J/mol ] & molar reduced chemical potential of component $k$\\
$\psi_{k}^{\measure}$ & [ J/kg ], [ J/mol ] & matter reduced chemical potential of component $k$\\
$\psi_{k,\,0}$ & [ J/mol ] & homogeneous molar reduced chemical potential of component $k$\\
$\psi_{k,\,0}^{\measure}$ & [ J/kg ], [ J/mol ] & homogeneous matter reduced chemical potential of component $k$\\
$c$ & [ mol/m\tsup{3} ] & total molar concentration\\
$c$ & [ mol/m\tsup{3} ] & short notation for the set $\{\arglist{c}{n}\}$ of molar concentrations (only in \secr{sec/NonUniform/Volume}) \\
$c_{k}$ & [ mol/m\tsup{3} ] & molar concentration of component $k$ \\
$F$ & [ J ] & total Helmholtz energy \\
$f$ & [ J/mol ] & specific Helmholtz energy per mole\\
$f_{0}$ & [ J/mol ] & homogeneous specific Helmholtz energy per mole\\
$f^{v}$ & [ J/m\tsup{3} ] & specific Helmholtz energy per unit of volume\\
$f^{\measure}_{0}$ & [ J/kg ], [ J/mol ] & homogeneous specific Helmholtz energy per unit of matter\\
$f^{v}_{0}$ & [ J/m\tsup{3} ] & homogeneous specific Helmholtz energy per unit of volume\\
$i,j,k$ & [ -- ] & component number\\
$m$ & [ kg ] & total mass \\
$m_{k}$ & [ kg ] & total mass of component $k$\\
$\measure$ & [ kg ], [ mol ] & total matter amount \\
$\measure_{k}$ & [ kg ], [ mol ] & total matter amount of component $k$\\
$\measure^{v}_{k}$ & [ kg/m\tsup{3} ], [ mol/m\tsup{3} ] & matter density of component $k$: amount of component $k$ per unit of volume\\
$n$ & [ -- ] & number of components\\
$p$ & [ Pa ] & scalar pressure \\
$p_{0}$ & [ Pa ] & homogeneous pressure \\
$\vR$ & [ m ] & position \\
$S$ & [ J/K ] & total entropy \\
$S$ & [ m\tsup{2} ] & the boundary surface \\
$s$ & [ J/(K mol) ] & specific entropy per mole\\
$s^{v}$ & [ J/(K m\tsup{3}) ] & specific entropy per unit of volume\\
$s^{\measure}_{0}$ & [ J/(K kg) ], [ J/(K mol) ] & homogeneous specific entropy per unit of matter\\
$s^{v}_{0}$ & [ J/(K m\tsup{3}) ] & homogeneous specific entropy per unit of volume\\
$T$ & [ K ] & temperature \\
$U$ & [ J ] & total internal energy \\
$u$ & [ J/mol ] & specific internal energy per mole\\
$u^{v}$ & [ J/m\tsup{3} ] & specific internal energy per unit of volume\\
$u^{\measure}_{0}$ & [ J/kg ], [ J/mol ] & homogeneous specific internal energy per unit of matter\\
$u^{v}_{0}$ & [ J/m\tsup{3} ] & homogeneous specific internal energy per unit of volume\\
$V$ & [ m\tsup{3} ] & total volume \\
$v$ & [ m\tsup{3}/mol ] & specific volume per mole\\
$v^{m}$ & [ m\tsup{3}/kg ] & specific volume per unit of mass\\
$v^{\measure}$ & [ m\tsup{3}/kg ], [ m\tsup{3}/mol ] & specific volume per unit of matter\\
$x_{\alpha}$ & [ m ] & cartesian coordinate in direction $\alpha$\\
\end{longtable}

%

\bibliographystyle{unsrt}

\end{document}